\pdfoutput=1

\documentclass[12pt]{iopart}

\usepackage{amscd,amssymb,a4}
\usepackage{graphicx,xspace}

\usepackage[left=3cm,top=3cm,right=3cm,bottom=3cm]{geometry}
\usepackage{color}

\begin{document}

\newcommand{\red}{\color{red}}
\newcommand{\blue}{\color{blue}}
\newcommand{\todo}[1]{\textbf{To do: #1}}
\newcommand{\be}{\begin{equation}}
\newcommand{\ee}{\end{equation}}
\newcommand{\beq}{\begin{equation}}
\newcommand{\eeq}{\end{equation}}
\newcommand{\bea}{\begin{eqnarray}}
\newcommand{\eea}{\end{eqnarray}}
\newcommand{\rar}{\rightarrow}
\newcommand{\lar}{\leftarrow}
\newcommand{\ra}{\right\rangle}
\newcommand{\la}{\left\langle }
\renewcommand{\d}{{\rm d }}
\newcommand{\m}{{\tilde m }}
\newcommand{\p}{\partial}
\newcommand{\nn}{\nonumber }

\newcommand{\fig}[2]{\includegraphics[width=#1]{./figures/#2}}
\newcommand{\Fig}[1]{\includegraphics[width=7cm]{./figures/#1}}
\newlength{\bilderlength}
\newcommand{\bilderscale}{0.35}
\newcommand{\storebilderscale}{\bilderscale}
\newcommand{\bilderskip}{\hspace*{0.8ex}}
\newcommand{\textdiagram}[1]{%
\renewcommand{\bilderscale}{0.2}%
\diagram{#1}\renewcommand{\bilderscale}{\storebilderscale}}
\newcommand{\vardiagram}[2]{%
\newcommand{\bilderscale}{#1}%
\diagram{#2}\renewcommand{\bilderscale}{\storebilderscale}}
\newcommand{\diagram}[1]{%
\settowidth{\bilderlength}{\bilderskip%
\includegraphics[scale=\bilderscale]{./figures/#1}\bilderskip}%
\parbox{\bilderlength}{\bilderskip%
\includegraphics[scale=\bilderscale]{./figures/#1}\bilderskip}}
\newcommand{\Diagram}[1]{%
\settowidth{\bilderlength}{%
\includegraphics[scale=\bilderscale]{./figures/#1}}%
\parbox{\bilderlength}{%
\includegraphics[scale=\bilderscale]{./figures/#1}}}
\bibliographystyle{KAY}

%


%








\newcommand{\atanh}
{\operatorname{atanh}}

\newcommand{\ArcTan}
{\operatorname{ArcTan}}

\newcommand{\ArcCoth}
{\operatorname{ArcCoth}}

\newcommand{\Erf}
{\operatorname{Erf}}

\newcommand{\Erfi}
{\operatorname{Erfi}}

\newcommand{\Ei}
{\operatorname{Ei}}

\newcommand{\sgn}{{\mathrm{sgn}}}
\def\be{\begin{equation}}
\def\ee{\end{equation}}

\def\bea{\begin{eqnarray}}
\def\eea{\end{eqnarray}}

\def\e{\epsilon}
\def\l{\lambda}
\def\d{\delta}
\def\o{\omega}
\def\cb{\bar{c}}
\def\Li{{\rm Li}}

\title[KPZ equation: crossover from droplet to flat initial conditions]{Crossover from droplet to flat initial conditions in the KPZ equation from the replica Bethe ansatz}

\author{Pierre Le Doussal}

\address{CNRS-Laboratoire de Physique Th\'eorique de l'Ecole Normale Sup\'erieure\\
24 rue Lhomond, 75231 Paris
Cedex-France}

%

%



\date{\today}

\begin{abstract}
We show how our previous result based on the replica Bethe ansatz for the Kardar Parisi Zhang (KPZ) equation with the "half-flat" 
initial condition leads to the Airy$_2$ to Airy$_1$ (i.e. GUE to GOE) universal crossover one-point height distribution
in the limit of large time. Equivalently, we obtain the distribution of the free energy of a long directed polymer (DP) in a random potential with one fixed endpoint and the other one on a half-line.
We then generalize to a DP when each endpoint is free on its own half-line. It amounts,
in the limit of large time, to obtain the distribution of the maximum of the transition process Airy$_{2\to 1}$ (minus a
half-parabola) on a half line. 
\end{abstract}

\maketitle

\newpage

\section{Introduction}

Recently there was a lot of progress in finding exact solutions to the one-dimensional noisy Kardar-Parisi-Zhang (KPZ) equation. This
equation \cite{KPZ} describes the growth of an interface, in the continuum, parameterized by its height field $h(x,t)$ at point $x$
and has numerous experimental realizations \cite{exp4,exp5}. 
The growth is generated by an additive space-time white noise, and the problem is to characterize the statistics of the height field 
as a function of time $t$. While the scaling exponents $h \sim t^{1/3}$, $x \sim t^{2/3}$ have been known for a while \cite{exponent}, the recent
focus is on the PDF (probability distribution function) of the height field. 
The KPZ problem can be mapped to the problem of a continous directed polymer (DP) in a quenched random potential, in such a way that $h(x,t)= \ln Z(x,t)$ is proportional to the free energy of the DP with endpoint at $x$ and length $t$. 

As was anticipated from exact solutions of discrete models which belong to the KPZ universality class,
such as the PNG growth model \cite{png,spohn2000,ferrari1}, the TASEP particle transport model \cite{spohnTASEP,ferrariAiry,Airy1TASEP} or discrete DP models
\cite{Johansson2000,spohn2000}, one expects only a few universal statistics at large time, depending on the type of initial condition.
Remarkably, the interface retains some memory of the initial condition, even at large time.

For the KPZ equation on the infinite line there are three main classes. The {\it droplet initial condition} (which corresponds to
a DP with two fixed endpoints) leads to height fluctuations governed at large time by the Tracy Widom (TW) distribution $F_2$,
the CDF (cumulative distribution function) of the largest eigenvalue of the GUE random matrix ensemble \cite{TW1994}. It was 
solved simultaneoulsy by two methods. The first route uses a limit from an ASEP model with weak asymetry
\cite{spohnKPZEdge} and has allowed for a rigorous derivation \cite{corwinDP,reviewCorwin}. The second route
\cite{we,dotsenko} uses methods of disordered systems, namely replica, and methods from integrable systems, namely the Bethe Ansatz. It works on the DP version and allows to calculate the integer moments of $Z=e^{h}$ from the known exact solution of the Lieb-Liniger delta Bose 
gas \cite{ll}.  Extracting the PDF for $h$ from the integer moments is, as yet, a non-rigorous step. For the droplet initial conditions, both methods obtained the CDF for all times $t$, in the form of a Fredholm determinant, nicely displaying convergence to $F_2$ as $t \to +\infty$. 

The second important class, {\it the flat initial condition}, was solved using the Replica Bethe Ansatz (RBA) \cite{we-flat,we-flatlong,flatnumerics} in the form of a Fredholm Pfaffian, valid at all times. At large time the CDF of the height converges to the TW distribution $F_1$ associated with the
GOE ensemble of random matrices. A rigorous derivation is only presently in progress \cite{Quastelflat}. In fact, remarkable developments have occured in the math community, from the study of the so-called $q$-TASEP and related models, which aim to produce many rigorous results as limit processes (e.g. as $q \to 1$) \cite{BorodinMacdo,BorodinQboson}. 

The RBA also allowed for the solution of the last important class, {\it the stationary KPZ} \cite{SasamotoStationary}, and of the KPZ equation on the half-line \cite{we-halfspace} which relates to the GSE random matrix ensemble. Note that all the above mentionned exact solutions arise due to an
emerging, and somewhat miraculous, Fredholm determinant or pfaffian structure, found to hold for arbitrary time. 
Another important aim is to use the RBA to derive systematically the large time asymptotics, even when the knowledge of the finite time result is
unavailable. This strategy was recently explored, leading to another set of results \cite{dotsenkoGOE,ps-2point,ps-npoint,dotsenko2pt,Spohn2ptnew,dotsenkoEndpoint,dotsenko2times}. The joint distribution of 
$h(x,t)$ at several space points was obtained \cite{ps-2point,ps-npoint,dotsenko2pt,Spohn2ptnew}. This is not strictly a "new" result since from
the PNG and TASEP models it was anticipated that the (scaled) many point statistics of $h(x,t)$ converge 
to the one of the Airy$_2$ process ${\cal A}_2(x)$ (minus a parabola) whose one-point CDF is given by $F_2$.
The Airy$_2$ process may be defined as the trajectory of the largest eigenvalue of the GUE Dyson Brownian motion
(for definition and review see e.g. \cite{ps-npoint,ferrariAiry}, see also \cite{corwinRG}). 
However, recovering this result within the RBA is a non-trivial and interesting result. 
Another breakthrough was the calculation within the RBA of the endpoint distribution of the DP \cite{dotsenkoEndpoint} directly for
infinite time, which was found to agree with the (simultaneous) result about the position of the maximum of
the Airy 2 process (minus a parabola) \cite{greg2,quastelendpoint,baikgreg,QuastelSupremumAi2}. 
In both cases, the corresponding finite time problem is unsolved and
seems very difficult. A genuinely new result is the recent calculation from RBA \cite{dotsenko2times} of the two-time distribution
for the KPZ equation at infinitely separated times. It is important to note that the manipulations leading directly
to the infinite time limit in the RBA involve a substantial amount of "guessing" which makes it even less rigorous.
However, from the point of view of heuristics it is a very interesting route to explore further. 

Besides these three main classes, one also expects three universal {\it crossover classes} (also called transition classes) 
with initial conditions which interpolate from one of the three classes at $x=-\infty$ to a distinct one at $x=+\infty$, see e.g. Fig. 4 in 
Ref. \cite{reviewCorwin}. The aim of the present paper is to study the transition from GUE to GOE statistics in the KPZ equation. 
This is realized for the so-called "half-flat" intial condition, which is flat to the left and droplet-like to the right. Interestingly, in 
Ref. \cite{we-flat,we-flatlong} we had already obtained the formula for the moments $\overline{Z^n}$ for the half-flat initial condition.
There we studied only the $x \to - \infty$ limit of this formula to solve the flat case for arbitrary time. This formula 
seems hard to analyze for arbitrary time, however here we consider its large time limit and obtain the PDF for the KPZ height in the form of a Fredholm determinant interpolating between the $F_2$ and $F_1$ distributions. We obtain a new closed formula for the Kernel and shows that it is equivalent, via some Airy function identities, with the one obtained in Appendix A of Ref.
\cite{BorodinAiry2to1} from a solution of the TASEP. The corresponding Airy process was defined and characterized 
there and called ${\cal A}_{2 \to 1}$.

Since the recipe for the large time limit seems to work, we extend the calculation to obtain a genuinely new result. We consider 
the DP problem in the situation where each endpoint is free on its own half-line. It can again be solved in terms of
Fredholm determinants, with new Kernels. Recast in terms of Airy processes, it amounts to obtain the distribution 
of the maximum of the transition process Airy$_{2\to 1}$ (minus a half-parabola) on a half line.

The outline of the paper is as follows. In Section \ref{sec:model} we recall the KPZ and DP models and their connection and define the
dimensionless units. In Section \label{sec:model} \ref{sec:known} we recall the known results for the droplet and the flat initial conditions. 
In Section \ref{sec:crossover} we explain what we aim to do in this paper, give elementary facts about the transition process Airy$_{2\to 1}$
and introduce the generating function. 
In Section \ref{sec:QM} we briefly recall the replica Bethe ansatz method and in Section \ref{sec:start},
the formula from Ref. \cite{we-flat,we-flatlong}. 
In Section \ref{sec:largetime} we consider the large time limit of this formula, and obtain the new form for the Kernel of the transition process.
In Section \ref{sec:equiv} we use identities between Airy functions to put it in a form which is then compared in Section \ref{sec:recover} 
to the previous results of Ref. \cite{BorodinAiry2to1}. Finally in Section \ref{sec:general}  we generalize the problem, obtain the new Kernels
for the so-called $LL$ and $LR$ problems and explain the connection to the extrema of the Airy$_{2\to 1}$ process. We conclude on open problems and give some details in the Appendices. 

\section{Model and dimensionless units} 
\label{sec:model}

\subsection{the KPZ equation}

Consider the standard 1D continuum KPZ growth equation for the height field $h(x,t)$:
\be \label{kpzeq}
\partial_t h = \nu \nabla^2 h + \frac{1}{2} \lambda_0 (\nabla h)^2 + \sqrt{D} ~ \eta 
\ee
in presence of the white noise $\overline{\eta(x,t) \eta(x',t')}=\delta(x-x') \delta(t-t')$.
We define the scales 
\bea
x_0 = (2 \nu)^3 \quad , \quad t_0 = 2 (2 \nu)^5 \quad , \quad \lambda_0  h_0 = 2 \nu
\eea 
and use them as units, i.e. we set $x \to x_0 x$, $t \to t_0 t$ and $h \to h_0 h$
and work from now on in the dimensionless units where the KPZ equation becomes:
\bea \label{kpzeq2}
&& \partial_t h =  \nabla^2 h + (\nabla h)^2 + \sqrt{2 \bar c} ~ \eta  \\
&& \bar c= D \lambda_0^2
\eea

\subsection{the directed polymer}

Consider now $Z(x,t|y,0)$ the partition function of the continuum directed polymer in the random potential
$- \sqrt{\bar c} ~ \eta(x,t)$ with fixed endpoints at $(x,t)$ and $(y,0)$, at temperature $T$:
\be \label{zdef} 
Z(x,t|y,0) = \int_{x(0)=y}^{x(t)=x}  Dx e^{- \frac{1}{T} \int_0^t d\tau [ \frac{1}{2}  (\frac{d x}{d\tau})^2  - \sqrt{\bar c} ~ \eta(x(\tau),\tau) ]}
\ee
As is well known it can be
mapped onto the KPZ equation with the correspondence:
\bea
\frac{\lambda_0}{2 \nu} h \equiv \ln Z \quad , \quad T=2 \nu 
\eea
Here and below, overbars denote averages over the white noise $\eta$.  
For both problems we define, as in Ref. \cite{we,dotsenko,we-flat} the dimensionless parameter $\sim t^{1/3}$:
\bea \label{lambdadef}
\lambda = \frac{1}{2} (\bar c^2 t/T^5)^{1/3} ,
\eea 
which measures the scale of the fluctuations of the DP free energy, i.e. of the KPZ height.
From now on we use the same units $x_0 = T^3$ and $t_0=2 T^5$ as above, and
in these dimensionless units the partition sum $Z=Z(x,t|y,0)$ is the solution of:
\bea \label{dp1} 
\partial_t Z = \nabla^2 Z + \sqrt{2 \bar c} ~ \eta Z
\eea 
with initial condition $Z(x=0,t|y,0)= \delta(x-y)$. In these
units the dimensionless parameter is $\lambda=(\bar c^2t/4)^{1/3}$. 

\subsection{Cole-Hopf mapping} 

The Cole-Hopf mapping solves the KPZ equation in terms of the DP partition sum, 
in the dimensionless units:
\be
e^{h(x,t)} = \int dy Z(x,t|y,0) e^{h(y,t=0)}.
\ee
which maps Eq. (\ref{kpzeq2}) into (\ref{dp1}). We will thus also adopt the notation:
\bea \label{drop0} 
h(x,t|y,0) = \ln Z(x,t|y,0)
\eea 
although it is somewhat improper since it requires a regularization near $t=0$ (see below). 

Below, when specified, we will often set $\bar c=1$ which amounts to a further change of units $x \to x/\bar c$ and $t \to t/\bar c^2$.

\section{Known results for the droplet and flat initial conditions}
\label{sec:known}

The droplet initial conditions for the KPZ equation is by definition the narrow wedge:
\bea
&& h_{drop}(x,t=0)= - w |x|  \quad , \quad w \to + \infty 
\eea 
and corresponds to the fixed endpoint initial condition $Z(x,t=0)=\delta(x)$ for the DP. 
More generally $h(x,t|y,0) \equiv h_{drop}(x-y,t)+ \ln(\frac{w}{2})$ given by (\ref{drop0}) corresponds to a sharp wedge
centered in $y$. Everywhere in this paper $\equiv$ means equivalent {\it in law}. The 
additive constant $\ln(\frac{w}{2})$ is necessary for regularization, but we will ignore below
all time-independent constants. 

It is known \cite{spohnKPZEdge,corwinDP,reviewCorwin,we,dotsenko} that at large time the one-point fluctuations of the height  
grow as $t^{1/3}$ and are governed by the 
GUE Tracy Widom (cumulative) distribution $F_2(s)$ as:
\bea \label{defxi}
&& h_{drop}(0,t) = \ln Z_{drop}(0,t) \simeq v_0 t + \bar c^{2/3} t^{1/3} \chi_2 \\
&&  {\rm Prob}(\chi_2 < s) = F_2(s) = {\rm Det}[I - P_0 K^s_{Ai} P_0] 
\eea  
where $F_2(s)$ is given by a Fredholm determinant with the Airy Kernel:
\bea \label{airyK} 
\fl && K_{Ai}^s(v_1,v_2)= K_{Ai}(v_1+s,v_2+s) \quad , \quad  K_{Ai}(v,v')= \int_{y>0} dy Ai(y+v) Ai(y+v')
\eea
and $P_0(v)=\theta(v)$ is the projector on $R^+$. 

More generally, for droplet initial conditions, the multi-point correlation is believed to converge \cite{ps-npoint,reviewCorwin,CorwinKPZensemble}
to the ones of the Airy$_2$ process ${\cal A}_2(u)$ 
\cite{spohn2000,ferrariAiry} with (in units \footnote{In several works, e.g. \cite{reviewCorwin,ps-npoint,QuastelSupremumAi2}, the
dimensionless KPZ equation is defined as (\ref{kpzeq}) with $\nu=\frac{1}{2}$, $\lambda_0=1$ and $D=1$. Compared to
our dimensionless problem with $\bar c=1$, this 
is equivalent to only a change of the time: let us denote the time $t'$ there, then $t'=2 t$ where $t$ denotes the time here.
We chose to conserve all notations and conventions of our previous works.} where $\bar c=1$):
\bea \label{h2} 
&& h(x,t) \simeq  v_0 t + t^{1/3} ( {\cal A}_2(u) - u^2 ) \quad , \quad u=\frac{x}{2 t^{2/3}} 
\eea 
where ${\cal A}_2(0) \equiv \chi_2$ and the process ${\cal A}_2(u)$ is stationary, i.e. statistically translationally 
invariant in $u$.

For the flat initial condition $h(x,t=0)=0$, it was found \cite{we-flat,we-flatlong} that \footnote{Here and above $v_0$
has a non-universal part depending on the regularization of the model at short scale. However, as detailed in
\cite{flatnumerics} if one considers $\ln (Z(t)/{\overline Z(t)})$ then it is universal with $v_0 = -\bar c^2/12$ in our units}:
\bea 
&& h_{flat}(0,t) = \ln Z_{flat}(0,t) = v_0 t + \bar c^{2/3} 2^{-2/3} t^{1/3} \chi_1 \\
&&  {\rm Prob}(\chi_1 < s) = F_1(s) = {\rm Det}[1 - P_0 B_s P_0] 
\eea 
where $F_1(s)$ is the GOE Tracy Widom (cumulative) distribution
which is expressed as a Fredholm determinant with 
the Kernel $B_s(v,v') = Ai(v+v'+s)$. Again it is believed that
in that case the joint distribution of the heights $\{ h_{flat}(x,t) \}_x$ are
governed by the so-called Airy$_1$ stationary process ${\cal A}_1(u)$ (switching back to units 
where $\bar c=1$):
\bea
&& h(x,t) \simeq  v_0 t + 2^{1/3}  t^{1/3} {\cal A}_1(2^{-2/3} u) 
\eea 
where ${\cal A}_1(0) = \frac{1}{2} \chi_1$. 
The Airy$_1$ process is related to the largest eigenvalue of the
GOE Dyson Brownian motion. For definition and normalizations
see e.g. Ref.  \cite{Airy1TASEP,ferrariAiry,QuastelSupremumAi2}.

Note that there is a connection between these results. Indeed from
the definition one expects, in the large time limit:
\bea
\fl && h_{flat}(x,t) = \ln \int dy ~ e^{h(x,t|y,0)} \equiv 
\ln \int dy ~ e^{h(y,t|x,0)} \simeq v_0 t +  t^{1/3} \max_u ( {\cal A}_{2}(u) - u^2) 
\eea 
where we have used that the sets $\{ h(x,t|y,0) \} \equiv \{ h(y,t|x,0) \}$ are statistically
equivalent and that, since height fluctuations grow as $t^{1/3}$, the 
integral is dominated by the maximum. 
Hence:
\bea
\max_u ( {\cal A}_2(u) - u^2)  = 2^{-2/3} \chi_1 = 2^{1/3}  {\cal A}_1(0)
\eea 
i.e. the maximum of the Airy$_2$ process minus a parabola is given by the
Airy$_1$ process at one point, as proved in \cite{QuastelSupremumAi2}.

\section{Aim of this paper: crossover from droplet to flat}

\label{sec:crossover}

\subsection{half-flat initial conditions and STS identity}

\begin{figure}[htpb]
  \centering
  \includegraphics[width=0.4\textwidth]{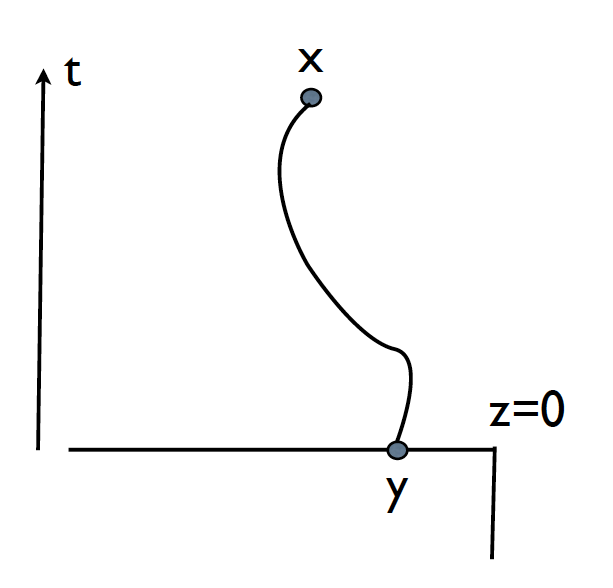}
  \caption{Half-flat initial conditions: one end-point of the DP is fixed at $x$, the other free on the half-line $y<z=0$. In addition
  there is a weight $e^{w y}$ which, from the STS symmetry, amounts to tilt the half-line (downward at $x=-\infty$ for $w>0$).}
\label{fig:fig1}
\end{figure}

Consider the double wedge initial condition for the KPZ field on the real axis:
\be
h(x,t=0)= w x ~ \theta(-x) - w' x ~ \theta(x) 
\label{wedgeic}
\ee
where $\theta(x)$ is the Heaviside step function. In this paper we 
focus on the limit $w' \to +\infty$. In terms of the DP it corresponds to a
(left) half-space problem with partition sum
\bea \label{hsdef} 
&& h_w(x,t)= \ln Z_{w}(x,t)  \\
&& Z_{w}(x,t) = \int_{-\infty}^{0} dy e^{w y} Z(x,t|y,0)\,,
\eea
with $Z_w(x,t=0)=\theta(-x) e^{w x}$. Hence for $w=0$ it can be seen as a "half-flat" initial condition, see Fig. \ref{fig:fig1}. 

Our aim here is to calculate the one-point probability distribution (PDF) of (minus) the free energy of the DP,
equivalently, of the height field of the KPZ equation. We write:
\bea \label{defxi}
&& -F = h = \ln Z = v_0 t + \lambda \xi_t \quad , \quad Z \equiv  Z_{w}(x,t)  \quad , \quad \lambda = (\bar c^2 t/4)^{1/3} 
\eea 
which defines $\xi_t$. 
This is done by calculating via the Bethe ansatz the integer moments of the partition function $\overline{Z^n}$.
Using the statistical tilt symmetry (STS) of the problem it is easy to show the exact relation for integer moments
(see e.g. \cite{we-flat} Appendix A):
\bea \label{stsrel}
\overline{Z^n_w(x,t)} = e^{-\frac{n x^2}{4 t}} \overline{Z^n_{w+\frac{x}{2 t}}(0,t)} \,.
\eea
equivalently:
\bea \label{sts1} 
\ln Z_w(x,t) + \frac{x^2}{4 t} \equiv_{\rm in ~ law} \ln Z_{w+\frac{x}{2 t}}(0,t) 
\eea 
Hence, up to a simple additive piece in the free energy, the problem depends only on the combination
$w + \frac{x}{2 t}$, i.e. changing the endpoint is the same as changing $w$. Since
we know that $h_0(-\infty,t) \equiv h_{flat}(0,t)$ and $h_{+\infty}(0,t) \equiv h_{drop}(0,t)$
there are thus two limits:
\bea
&& w + \frac{x}{2 t} \to + \infty  \quad , \quad h_{w+\frac{x}{2 t}}(0,t) \equiv h_{drop}(0,t)   \\
&& w + \frac{x}{2 t} \to - \infty   \quad , \quad h_{w+\frac{x}{2 t}}(0,t) \equiv h_{flat}(0,t)
\eea 
hence the present problem extrapolates from droplet to flat initial conditions. 

\subsection{relation to the ${\cal A}_{2\to 1}$ transition process}

Although this crossover has not yet been fully studied for the KPZ equation, 
it has been studied in Ref. \cite{BorodinAiry2to1} for the TASEP with an initial condition where particles are
placed at the even integers, the equivalent of the "half-flat" initial condition. There the transition process ${\cal A}_{2 \to 1}(u)$
was defined. It enjoys the two limits \footnote{There we follow Ref.
\cite{QuastelSupremumAi2}.} 
\bea
&& \lim_{u_1 \to + \infty} {\cal A}_{2 \to 1}(u + u_1) = 2^{1/3} {\cal A}_1 (2^{-2/3} u) \\
&& \lim_{u_1 \to - \infty} {\cal A}_{2 \to 1}(u + u_1) = {\cal A}_2(u)
\eea 
It was later shown \cite{QuastelSupremumAi2} that this transition process satisfies:
\bea
\max_{v < u} ({\cal A}_2(v)-v^2)  + \min(0,u)^2 = {\cal A}_{2 \to 1}(u) 
\eea 

It is then clear that the crossover studied here in the KPZ equation should be related to
this transition process in the limit of infinite time. Indeed, let us generalize slightly our notations and define:
\bea
&& h^L_{zw}(x,t) = \ln \int_{y < z} dy ~ e^{h(x,t|y,0)+w y} 
\eea 
so that $h_w(x,t)=h^L_{0w}(x,t)$. Then we again expect that in the large time
limit this solution is related to the maximum of the droplet solution on
a half-line \footnote{for convenience we drop here and below the constant $v_0 t$, which is
easily restored.} 
\bea
\fl && h^L_{z0}(x,t) = \ln \int_{y < z} dy e^{h(x,t|y,0)} \equiv 
\ln \int_{y < z} dy e^{h(y,t|x,0)} \\
\fl && ~~~~~~~~~~~~~~~~~~~~~~~~~~~~~~
~~~~~~~~ \simeq_{t \to \infty} \max_{y<z} h(y,t|x,0) \equiv \max_{y<z-x} h_{drop}(y,t) \nn
\eea
From (\ref{h2}) one then sees that (in units $\bar c=1$):
\bea
\fl && ~~~~~ h^L_{z}(x,t) \simeq_{t \to \infty}  t^{1/3} \max_{v < \frac{z-x}{2 t^{2/3}}} ( {\cal A}_2(v) - v^2 ) 
= t^{1/3} [ {\cal A}_{2 \to 1}(u) - \min(0,u)^2 ]_{u=\frac{z-x}{2 t^{2/3}}} \label{hLAiry} 
\eea 

More generally:
\bea
\fl &&~~~~ h^L_{zw}(x,t) \simeq_{t \to \infty} \max_{y<z} ( h(y,t|x,0) + w y) \equiv w x + \max_{y<z-x} ( h_{drop}(y,t) + w y) \\
\fl && ~~~~~~~~~~~~~~~ = w x + t^{1/3}  \max_{2 t^{2/3} v < z-x} ( - v^2 +  {\cal A}_2(v)  + 2^{2/3} \tilde w \times 2 v ) \\
\fl && ~~~~~~~~~~~~~~~ \equiv - \frac{x^2}{4 t} + t (w + \frac{x}{2 t})^2 + \max_{v < u} ( - v^2 +  {\cal A}_2(v)  ) \\
\fl && ~~~~~~~~~~~~~~~ = - \frac{x^2}{4 t} + t (w + \frac{x}{2 t})^2 + t^{1/3} [ {\cal A}_{2 \to 1}(u) - \min(0,u)^2 ]_{u=- t^{1/3} (w + \frac{x-z}{2 t})} \label{resa21} 
\eea 
where we have defined the scaled variable $\tilde w = \lambda w$ and used the stationarity of ${\cal A}_2$ 
to shift its argument. We also note that the last line can be rewritten as:
\bea
h^L_{zw}(x,t) \simeq_{t \to \infty} = h^{L0}_{zw}(x,t) +  t^{1/3}  {\cal A}_{2 \to 1}(u) 
\eea 
where $h^{(L0)}_{zw}(x,t)  = \max_{y<z} (-\frac{(y-x)^2}{4 t} + w y)$ is the large time solution of the KPZ equation with the
same initial conditions, in the absence of noise. 
This shows why the term $- \min(0,u)^2$ has been included in the definition of the transition Airy process
chosen in Ref. \cite{BorodinAiry2to1}. 

%

\subsection{generating function and its large time limit}
\label{sec:generate}

To later extract the PDF from the moments we now introduce the generating function, as in our previous works
\cite{we,we-flat,we-flatlong}:
\bea \label{defg} 
\fl &&  ~~~~~~ g_\lambda(s) = \overline{e^{- e^{- \lambda s} Z_{w}(x,t)}} = 1 + \sum_{n=1}^\infty \frac{(- e^{- \lambda s})^n}{n!} \overline{Z_{w}(x,t)^n} = 
\overline{ \exp( - e^{- \lambda (s - \xi)} ) } 
\eea
Once $g_\lambda(s)$ is known, the PDF of (minus) the rescaled free energy, $P(\xi)$, 
at large time (i.e. $\lambda  \to \infty$) is immediately extracted as:
\be \label{conv}
g_{\infty}(s) =  \lim_{\lambda \to \infty} g_\lambda(s) = \overline{ \theta(s-\xi) } = {\rm Prob}(\xi < s)  \,. 
\ee

Here we will calculate the generating function $g_{+\infty}(s)$ for the half-flat initial condition using the replica Bethe Ansatz. One will check that it
does reproduce correctly the two limits:
\bea
&& w + \frac{x}{2 t} \to + \infty  \quad , \quad  g_\infty(s) = \lim_{t \to \infty} {\rm Prob}(\xi_t < s) = F_2(2^{-2/3} s)    \\
&& w + \frac{x}{2 t} \to - \infty   \quad , \quad  g_\infty(s) = \lim_{t \to \infty} {\rm Prob}(\xi_t < s) = F_1(s) 
\eea 
in terms of the scaled random variable $\xi_t$ defined in (\ref{defxi}), where $F_1(s)$ and $F_2(s)$ are respectively the GOE and GUE Tracy Widom distributions.

\section{Quantum mechanics and Bethe Ansatz}
\label{sec:QM} 

The calculation of the $n$-th integer moment of the DP partition sum can be expressed \cite{kardareplica,bb-00} using the eigenstates
$\Psi_{\mu}$ and eigenenergies $E_\mu$ of the {\it attractive} Lieb-Liniger Hamiltonian for $n$ bosons \cite{ll}: 
\be
H_n = -\sum_{\alpha=1}^n \frac{\partial^2}{\partial {x_\alpha^2}}  - 2 \bar c \sum_{1 \leq  \alpha< \beta \leq n} \delta(x_\alpha - x_\beta).
\label{LL}
\ee
namely \footnote{for convenience we take the complex conjugate, since the total expression is real
it is immaterial} \cite{we-flat}:
\bea
\overline{Z_{w}(x,t)^n}&=&
\sum_\mu \frac{\Psi^*_\mu(x,..x)}{||\mu ||^2} e^{-t E_\mu}  
 \Big(\prod_{\alpha=1}^n \int_{-\infty}^0 dy_\alpha e^{w y_\alpha} \Big) \Psi_\mu(y_1\dots y_n)\,.
 \label{sumf}
\eea
These eigenstates are known from the Bethe ansatz \cite{ll}. They are parameterized by a set of rapidities
$\mu \equiv \{ \lambda_1,..\lambda_n\}$ which are solution of a set of coupled equations, the Bethe equations
(see below). They take the (un-normalized) form (totally symmetric in the $x_\alpha$): 
\be \label{def1}\fl
\Psi_\mu(x_1,..x_n) =  \sum_P A_P \prod_{j=1}^n e^{i \sum_{\alpha=1}^n \l_{P_\alpha} x_\alpha} \, , \quad 
A_P=\prod_{n \geq \beta > \alpha \geq 1} \Big(1 + \frac{i \bar c ~\sgn(x_\beta - x_\alpha)}{\lambda_{P_\beta} - \lambda_{P_\alpha}}\Big)\,.
\ee
where the sum runs over all $n!$ permutations $P$ of the rapidities $\l_j$. The corresponding eigenenergies are
$E_\mu=\sum_{\alpha=1}^n \lambda_\alpha^2$. In the formula (\ref{sumf}) we also need: 
\be 
\Psi^*_\mu(x,..x) = n! e^{-  i x \sum_\alpha \lambda_\alpha} \,.
\label{Psixeq}
\ee 
Before discussing the last two ingredients in (\ref{sumf}), i.e. the the norms $||\mu||^2$ and the half-space 
integrals of the Bethe eigenfunctions, let us recall the spectrum of $H_n$ in the limit of infinite system size,
i.e. the rapidities solution to the Bethe equations \cite{m-65} . 
A general eigenstate is built by partitioning the $n$ particles into a set of $n_s \leq n$ bound states called {\it strings} 
formed by $m_j \geq 1$ particles with $n=\sum_{j=1}^{n_s} m_j$. 
The rapidities associated to these states are written as
\be\label{stringsol}
\l^{j, a}=k_j +\frac{i\cb}2(m_j+1-2a) 
\ee 
Here, $a = 1,...,m_j$ labels the rapidities within the string $j=1,\dots n_s$. Inserting 
these in (\ref{def1}) leads to the Bethe eigenstates of the infinite system, and their corresponding
eigen-energies:
\be \label{en} 
E_\mu= \sum_{j=1}^{n_s} m_j k_j^2-\frac{\cb^2}{12} m_j(m_j^2-1).
\ee 

For now on, we use units where $\bar c=1$. The calculation of the norms is involved. The result however is simple: in the 
large system size $L$ limit it can be written as \cite{cc-07}:
\bea
&& \fl  \frac{1}{||\mu ||^2} = \frac{1}{n!  L^{n_s} } \prod_{1\leq i<j\leq n_s} 
\Phi_{k_i,m_i,k_j,m_j}  \prod_{j=1}^{n_s} \frac1{m_j^{2}}  \, , \quad  \Phi_{k_i,m_i,k_j,m_j}=\frac{4(k_i-k_j)^2 +(m_i-m_j)^2}{4(k_i-k_j)^2 +(m_i+m_j)^2}
 \nn \\
&& \fl \label{norm}
\eea

Consider now the integral in (\ref{sumf}) over the negative half-space. Since the wave functions are totally symmetric 
in their arguments, it can be 
performed on the sector $y_1<y_2<..y_n<0$. Using that:
\begin{eqnarray} \label{Gla} 
\fl &&  G_{\lambda} = \int_{-\infty< y_1<y_2< ..<y_n < 0} 
dy_1.. dy_n e^{\sum_{\alpha=1}^n (w+ i \lambda_\alpha) y_\alpha }  = \prod_{j=1}^{n} \frac{1}{ j w + i \lambda_1 + .. + i \lambda_j }\,,
 \nonumber 
\end{eqnarray}
a "miracle" occurs upon performing the summation over the permutations, leading to the factorized form \cite{we-flat,we-flatlong}:
\bea
&& \Big(\prod_{\alpha=1}^n \int_{-\infty}^0 dy_\alpha e^{w y_\alpha} \Big) \Psi_\mu(y_1\dots y_n) = \sum_P A_P G_{P \lambda} \nn \\
&& = \frac{n!}{ \prod_{\alpha=1}^n (w+ i \lambda_\alpha)}  \prod_{1 \leq \alpha < \beta \leq n} \frac{2 w + i \lambda_\alpha + i \lambda_\beta - 1}{ 2 w+ i \lambda_\alpha + i \lambda_\beta} . \label{miracle} 
\eea

Let us now explain a point which was implicit in \cite{we-flat,we-flatlong}. 
Note that, strictly, (\ref{Gla}) is valid only for $Re(j w + i \lambda_1 + .. + i \lambda_j)>0$ for all $j$. Hence (\ref{miracle}), which involves all
$G_{P \lambda}$, is
valid a priori when all $\lambda_j$ are real and $w>0$. However, one easily sees that its validity is more general. An important 
property of $\Psi_\mu$ for a string state is that $A_P = 0$ (for $x_1<..<x_n$) unless the imaginary parts of the rapidities belonging to a given
string are in increasing order. For instance for $n_2=2,m_1=2,m_2=3$, $A_P$ is non-zero only for $(\lambda_1,..\lambda_5)$ obtained from
$(k_1-\frac{i \bar c}{2},k_1+\frac{i \bar c}{2},k_2-\frac{3 i \bar c}{2},k_2-\frac{i \bar c}{2},k_2+\frac{i \bar c}{2},k_2+\frac{3 i \bar c}{2})$ 
by a permutation of $S_5$ which respects the order of the imaginary parts inside each string (e.g. $k_1-\frac{i \bar c}{2}$ must always appear before $k_1+\frac{i \bar c}{2}$). From this it is easy to see that the condition $Re(j w + i \lambda_{P_1} + .. + i \lambda_{P_j})>0$ for all $j$
is satisfied for all terms with non zero $A_P$. Physically it just expresses the fact that the bound states have a convergent integral
over space, and $w>0$ is needed only to make the integral over the center of mass convergent. 

\section{Previous result: starting formula for the generating function} \label{sec:starting2} 
\label{sec:start} 

Let us recall the derivation in Ref. \cite{we-flat,we-flatlong}, as it will be needed for further generalizations. 
From the ingredients (\ref{Psixeq}), (\ref{en}), (\ref{norm}), (\ref{miracle}) we can now perform the summation $\sum_\mu$ over the eigenstates in (\ref{sumf}).
It factors into a sum over the string variables $k_j,m_j$. One shows that in the infinite system 
the string momenta $m_j k_j$ are quantized as free particles, hence we can replace $\sum_{k_j} \to m_j L \int \frac{dk_j}{2 \pi}$,
all factors $L$ cancel with the norms. 
One obtains the moments $\overline{Z^n}$ as a sum $(m_1,\dots m_{n_s})_n$ over all the partitioning of $n$:
\bea \label{startZ} 
\fl && \overline{Z^n} = \sum_{n_s=1}^n \frac{n! 2^n}{n_s!} \sum_{(m_1,..m_{n_s})_n} 
\prod_{j=1}^{n_s} \int \frac{dk_j}{2 \pi} \frac{1}{m_j} e^{m_j^3  \frac{t}{12}- m_j k_j^2 t  - i x m_j k_j } 
\Phi[k,m] S^w[k,m] D^w[k,m]  
\eea
together with the generating function as an expansion in the number of strings:
\begin{eqnarray}  
&&  g_\lambda(s) =  1 +  \sum_{n_s=1}^\infty \frac1{n_s!}  Z(n_s,s)  \\
&& \fl 
 Z(n_s,s) = \label{start} \\
 && \fl \quad \sum_{m_1,\dots m_{n_s}=1}^\infty 
 \prod_{j=1}^{n_s} \frac{2^{m_j}}{m_j} \int \frac{dk_j}{2 \pi} 
S^w_{m_j,k_j}  e^{m_j^3  \frac{t}{12}- m_j k_j^2 t - \lambda m_j s - i x m_j k_j} 
\prod_{1 \leq i < j \leq n_s} \tilde D^w_{m_i,k_i,m_j,k_j}  \nn
\end{eqnarray}
In $g_\lambda(s)$ the summations over the $m_j$ are free. The factors $S^w$ and $D^w$ are obtained 
by inserting the string rapidities (\ref{stringsol}) into (\ref{miracle}). They read: 
\bea \label{sw}
&&  S^w_{m,k} = \frac{(-1)^m \Gamma(z)}{\Gamma(z + m)}  \quad , \quad z= 2 i k + 2 w.
\eea
and
\bea \label{dw}
\fl  D^w_{m_1,k_1,m_2,k_2}& =&
\frac{\Gamma \left(1-z -\frac{m_1+m_2}{2}\right) \Gamma \left(1-z +\frac{m_1+m_2}{2}\right) }{
\Gamma \left(1-z +\frac{m_1-m_2}{2}\right) \Gamma \left(1-z -\frac{m_1-m_2}{2}\right)}
\\ \fl
&  =& (-1)^{m_2} \frac{\Gamma(1- z + \frac{m_1+m_2}{2}) \Gamma(z + \frac{m_1-m_2}{2}) }{
\Gamma(1- z + \frac{m_1-m_2}{2}) \Gamma(z + \frac{m_1+m_2}{2})} \quad , \quad z=  i k_1+i k_2 + 2 w\,. \label{dw2}
 \eea
and in (\ref{startZ}), (\ref{start}) we have defined the notations:
\bea
&& \tilde D^w_{m_i,k_i,m_j,k_j} =  D^w_{m_i,k_i,m_j,k_j} \Phi_{k_i,m_i,k_j,m_j} \quad , \quad S^w[k,m]=  \prod_{j=1}^{n_s} S^w_{m_j,k_j}  \\
&& D^w[k,m]= \prod_{1 \leq i < j \leq n_s} D^w_{m_i,k_i,m_j,k_j} \quad, \quad \Phi[k,m]= \prod_{1 \leq i < j \leq n_s}  \Phi_{k_i,m_i,k_j,m_j} 
\eea 
involving the factor (\ref{norm}) coming from the norm. Note that we have also performed the
shift $Z \to Z e^{\bar c^2 t/12}$ which yields the factor $-\bar c^2/12$ in $v_0$ in Section \ref{sec:known}. 

Two remarks are in order:

(i) The STS relation (\ref{stsrel}) between moments can be retrieved by performing the change of
integration variable in the integral (\ref{start}):
\bea
i k_j \to i k_j + \frac{x}{2 t}  
\eea 
resulting in the global shift $Z(n_s,s)|_{x,w}  \to Z(n_s,s)|_{0,w'} e^{- \frac{x^2}{4 t} \sum_j m_j }$ with a new value $w \to w'=w + \frac{x}{2 t}$.
Since we know that this STS relation holds this shift (followed by shifting the integration contour back to the real axis) must be
legitimate provided $w'=w + \frac{x}{2 t}>0$ since this is the assumption used to derive (\ref{start}).

(ii) while the formula for the moments $\overline{Z^n}$ is well defined, because of the exponential cubic divergence of the series, the formula 
(\ref{start})  should be taken in some analytical continuation sense, i.e. it is valid
as a formal series in $t$. One way to do that, as discussed in \cite{we,dotsenko} and below, is to
use the {\it Airy trick}, valid for ${\rm Re}(z)>0$:
\be  \label{airytrick} 
 \int_{-\infty}^\infty dy Ai(y) e^{y z} = e^{z^3/3}\,.
\ee
the summations over $m$ being then carried later at fixed $y$ and usually convergent. 

\section{Rescaling and Airy trick} 
\label{sec:rescaling} 

Let us first perform some rescaling and rearrangement of our starting expression, to make
easier the large time limit in the next Section. In the $\bar c=1$ dimensionless units 
one has $t=4 \lambda^3$. One can the define the scaled position and slope:
\bea
w = \frac{\tilde w}{\lambda}=\frac{\tilde w}{(t/4)^{1/3}} \quad , \quad x=\lambda^2 \tilde x = (t/4)^{2/3} \tilde x
\eea 
such that $\tilde w$ and $\tilde x$ will be kept finite in the large time limit. One then perform the change $k_j \to k_j/\lambda$ in the integral. 
We get:
\bea
&& \fl 
 Z(n_s,s) =   
 \prod_{j=1}^{n_s} \sum_{m_j=1}^{+\infty}  \frac{2^{m_j}}{\lambda m_j} \int \frac{dk_j}{2 \pi}  \frac{(-1)^{m_j} \Gamma(\frac{2 i k_j + 2 \tilde w}{\lambda})}{\Gamma(\frac{2 i k_j + 2 \tilde w}{\lambda} + m_j)} 
 e^{\frac{1}{3} \lambda^3 m_j^3  - 4 \lambda m_j k_j^2  - \lambda m_j s - i  \lambda m_j k_j \tilde x}  \label{startresc} \\
&& \fl \times \prod_{1 \leq i < j \leq n_s} D^w_{m_i,k_i/\lambda,m_j,k_j/\lambda} \frac{4(k_i-k_j)^2 +\lambda^2 (m_i-m_j)^2}{4(k_i-k_j)^2 +\lambda^2 (m_i+m_j)^2}.\nn
\eea
An equivalent expression is obtained using the Airy trick and the double Cauchy identity:
\bea
\fl && \prod_{1 \leq i < j \leq n_s}  \frac{4(k_i-k_j)^2 +\lambda^2 (m_i-m_j)^2}{4(k_i-k_j)^2 +\lambda^2 (m_i+m_j)^2} =  det[ \frac{1}{ 2 i (k_i - k_j) + \lambda m_i + \lambda m_j }]_{n_s \times n_s} \prod_{j=1}^{n_s} (2 \lambda m_j) \nn
\eea 
followed by the shift $y_j \to y_j + s+ 4 k_j^2$, which leads to: 
\bea 
&& \fl 
 Z(n_s,s) =
 \prod_{j=1}^{n_s} \sum_{m_j=1}^{+\infty} 2^{m_j+1} \int \frac{dk_j}{2 \pi} dy_j \frac{(-1)^{m_j} \Gamma(\frac{2 i k_j + 2 \tilde w}{\lambda})}{\Gamma(\frac{2 i k_j + 2 \tilde w}{\lambda} + m_j)} Ai(y_j+ 4 k_j^2 + s)
 e^{\lambda m_j (y_j  - i  k_j \tilde x) }  \nn \\
&&  \times \prod_{1 \leq i < j \leq n_s} D^w_{m_i,k_i/\lambda,m_j,k_j/\lambda} \times det[ \frac{1}{ 2 i (k_i - k_j) + \lambda m_i + \lambda m_j }]_{n_s \times n_s}  \label{2} 
\eea 
Note that we know from the STS relation (\ref{sts1}) that $g(s)$ only depends on the following combination of variables:
\bea \label{gsts} 
g_\lambda(s;w,x) = \tilde g_\lambda(s + \frac{\tilde{x}^2}{16} , \tilde w + \frac{\tilde x}{8} ) 
\eea 

\section{Large time limit and Fredholm determinant form}
\label{sec:largetime} 

We now study the limit of large time, i.e. $\lambda \to +\infty$.  We will now {\it assume} that in this limit we can set the complicated factor $D^w_{m_i,k_i/\lambda,m_j,k_j/\lambda} \to 1$, which we do from now on. 
This appears to be true {\it a posteriori} from the result we will obtain. An attempt at a justification
is discussed in \ref{sec:large}. 

Let us evaluate the resulting expression setting $D^w_{m_i,k_i/\lambda,m_j,k_j/\lambda} \to 1$ in (\ref{2}), keeping for now arbitrary $\lambda$ and
keeping in mind that ultimately we will be interested in the limit $\lambda \to + \infty$. Using the expression of a determinant as a sum of permutations $\sigma \in S_{n_s}$ of signature $(-1)^\sigma$, ${\rm det} M = \sum_\sigma (-1)^\sigma \prod_{j=1}^{n_s} M_{j,\sigma(j)}$, and the reexponentiation
formula $\frac{1}{a} = \int_{v>0} e^{-a v}$ we introduce $n_s$ auxiliary variables $v_j$ and rewrite:
\bea
\fl && 
 Z(n_s,s) = \sum_{\sigma} (-1)^\sigma
 \prod_{j=1}^{n_s} \sum_{m_j=1}^{+\infty} 2^{m_j+1} \int_{v_j>0} \int \frac{dk_j}{2 \pi} dy_j \frac{(-1)^{m_j} \Gamma(\frac{2 i k_j + 2 \tilde w}{\lambda})}{\Gamma(\frac{2 i k_j + 2 \tilde w}{\lambda} + m_j)} Ai(y_j+ 4 k_j^2 + s)
   \nn \\ 
\fl && \times e^{\lambda m_j (y_j  - i  k_j \tilde x) - 2 i (k_j - k_{\sigma(j)}) v_j - v_j (m_j + m_{\sigma(j)})}
\label{3} 
\eea 
We can now use $\sum_j v_j a_{\sigma(j)} = \sum_j v_{\sigma^{-1}(j)} a_{j}$ and relabel the sum over permutations as $\sigma \to \sigma^{-1}$
to obtain:
\bea
\fl && 
 Z(n_s,s) = \sum_{\sigma} (-1)^\sigma
 \prod_{j=1}^{n_s} \sum_{m_j=1}^{+\infty} 2^{m_j+1} \int_{v_j>0}  \int \frac{dk_j}{2 \pi} dy_j \frac{(-1)^{m_j} \Gamma(\frac{2 i k_j + 2 \tilde w}{\lambda})}{\Gamma(\frac{2 i k_j + 2 \tilde w}{\lambda} + m_j)} Ai(y_j+ 4 k_j^2 + s)
   \nn \\ 
\fl && \times e^{\lambda m_j (y_j  - i  k_j \tilde x) - 2 i  k_j (v_j - v_{\sigma(j)})  - m_j ( v_j + v_{\sigma(j)})}
\label{3} 
\eea 
where now we can further shift the variable $y \to y + v_j + v_{\sigma(j)} + i k_j \tilde x$ using the identity (for $\lambda>0$):
\bea
\int_{-\infty}^{+\infty} dy Ai(y) e^{\lambda (y - i k x)} = \int_{-\infty}^{+\infty} dy Ai(y+ i k x) e^{\lambda y} 
\eea 
Under this form it clearly appears that $Z(n_s,s)$ has a determinantal form:
\bea
Z(n_s,s) = \prod_{j=1}^{n_s} \int_{v_j>0}  {\rm det} M(v_i,v_j)|_{n_s \times n_s} 
\eea 
with the Kernel:
\bea
\fl && M(v_1,v_2) =   \int \frac{dk}{2 \pi} dy 
Ai(y + 4 k^2 + i k \tilde x + v_1 + v_2 + s) e^{-2 i k (v_1-v_2)} \phi_\lambda(k,y) \label{M} \\
\fl && \phi_\lambda(k,y) = 
\sum_{m=1}^\infty \frac{(-1)^{m} 2^{m+1} \Gamma(\frac{2 i k + 2 \tilde w}{\lambda})}{\Gamma(\frac{2 i k + 2 \tilde w}{\lambda} + m)} e^{\lambda m y} 
\eea 
which we now study in the large $\lambda$ limit. We show that:
\bea \label{lim1} 
\phi_{+\infty}(k,y) = - 2 \theta(y) - \frac{1}{i k + \tilde w} \delta(y) 
\eea 
There are several ways to do that, the simplest is to use the Mellin-Barnes (MB) identity:
\bea \label{MB} 
\sum_{m=1}^{+\infty} (-1)^{m} f(m) = \frac{-1}{2 i} \int_{C}  ds \frac{1}{\sin \pi s} f(s) 
\eea 
where $C_j=a+i r$, $0<a<1$, $r \in ]-\infty,\infty[$. It requires being able to deform the contour and close it
around the positive real axis, picking up the residues of the inverse sine function. Clearly the
conditions for that are (i) $f(s)$ has no pole for $Re(s) \geq a$ (ii) $f(s)$ does not grow too fast at large $Re(s)>0$. The first condition
is necessary for us not to "miss" any pole: in fact if there are such poles in $f(s)$ they can just be added to the formula.
Here both conditions are clearly satisfied. Hence we obtain:
\bea
\phi_\lambda(k,y) = \frac{-1}{2 i} \int_{C}  ds \frac{1}{\sin \pi s}  
2^{s+1} \frac{\Gamma(\frac{2 i k + 2 \tilde w}{\lambda})}{\Gamma(\frac{2 i k + 2 \tilde w}{\lambda} + s)} e^{\lambda s y} 
\eea 
rescaling $s_j \to s_j/\lambda$, we obtain, in the limit $\lambda \to + \infty$:
\bea
\phi_{+\infty}(k,y) =  \int_{C}  \frac{- ds}{2 i \pi s}  \frac{2 i k + 2 \tilde w + s}{i k + \tilde w} e^{s y} 
\eea 
where now the contours $C_j$ are the same as above, but with $a=0^+$. We used that
$\Gamma(x/\lambda) \sim \lambda/x$ at large $\lambda$. We can now perform the integrals over the $s_j$, using:
\bea
 \int_{C}   \frac{- ds}{2 i \pi s}   e^{s y } = - \theta(y) \quad , \quad  \int_{C}   \frac{- ds}{2 i \pi}   e^{s y} = - \delta(y)
\eea 
and we obtain (\ref{lim1}). Note that "undoing" the MB trick one sees that at large $\lambda$:
\bea
\fl && \phi_\lambda(k,y)  \simeq \sum_{m=1}^\infty (-1)^{m} \frac{2 i k + 2 \tilde w + \lambda m}{i k + \tilde w} e^{\lambda m y} \\
&& =  ( 2 + \frac{1}{i k + \tilde w}  \partial_y)  \frac{e^{\lambda y}}{1+ e^{\lambda y}} \to_{\lambda \to +\infty} - 2 \theta(y) - \frac{1}{i k + \tilde w} \delta(y) 
\eea 
re-obtaining the same result via a direct summation over $m$ \footnote{Note that summation over $m$ of
for $\phi_\lambda(k,y)$ is also possible at any $\lambda$ leading to an hypergeometric function, see Appendix E.4 of 
\cite{we-flatlong}.} 

Inserting (\ref{lim1}) into the Kernel $M$ we finally obtain the large $\lambda$ limit:
\bea
Z(n_s,s) = \prod_{j=1}^{n_s} (-1)^{n_s} \int_{v_j>0}  {\rm det} K(v_i,v_j)|_{n_s \times n_s} 
\eea 
Hence the generating function takes the form of a Fredholm determinant:
\bea \label{fd1} 
g_{\infty}(s) = {\rm Det}[I - {\cal K}] \quad , \quad {\cal K}(v_1,v_2) = \theta(v_1) \theta(v_2) K(v_1,v_2) 
\eea 
with the Kernel:
\bea  \label{K0} 
\fl && K(v_1,v_2) =   \int \frac{dk}{2 \pi} dy ( 2 \theta(y) + \frac{\delta(y) }{i k + \tilde w} )
 Ai(y + 4 k^2 + i k \tilde x + v_1 + v_2 + s) e^{-2 i k (v_1-v_2)} 
\eea 
which is our main result, with here $\tilde w>0$, and we recall $\tilde x=x/\lambda^2$ and $\tilde w=\lambda w$. 
The STS symmetry can be checked by performing a shift $i k \to i k + \frac{\tilde x}{8}$ and bringing back the contour to the real axis.
This shifts $w \to w + \frac{\tilde x}{8}$ and $s \to s + \frac{\tilde x^2}{16}$ in agreement with (\ref{gsts})
\cite{footnote1}. Hence $\tilde g_{\infty}$ in that formula is also given by (\ref{fd1}) and the Kernel $K$ but with $\tilde x$ set to 0. 

The GUE limit is easy to check on this form for $K$. Setting $\tilde x=0$ and 
$\tilde w \to + \infty$ the second part of $K$ vanishes and one recovers exactly the
GUE Kernel in the form given in the Eq. (26) of \cite{we} (after the change $k \to k/2$). 

To check the GOE limit is more delicate. One can take $w=0^+$ and let $x \to - \infty$. 
In that limit it turns out that one can replace:
\bea
\frac{1}{i k + 0^+} \to 2 \pi \delta(k) 
\eea 
(a related property was noted in \cite{we-flatlong}), and that the first term vanishes. 
Hence $K \to B_s$ the kernel of the $F_1$ distribution.

We now give an equivalent form of the Kernel where the limits can be conveniently studied. 

\section{Equivalent forms for the Kernel}
\label{sec:equiv}

We now display three useful identity involving Airy functions. The first one is:
\bea \label{id1} 
\fl &&  2 \int \frac{dk}{2 \pi} Ai(4 k^2 + a+b+ i k \tilde x) e^{2 i k (b-a)} = 2^{-\frac{1}{3}}  Ai\big(2^{\frac{1}{3}} (a+\frac{\tilde x^2}{32})\big) Ai\big(2^{\frac{1}{3}} (b+\frac{\tilde x^2}{32})\big) e^{\frac{\tilde x}{4} (b-a)} \nn \\
\fl && 
\eea 
where $a,b,k,\tilde x$ are real numbers. This identity is well known in the case $\tilde x=0$ \cite{airy}. Starting from the identity for $\tilde x=0$ we can
shift $i k \to i k - \frac{\tilde x}{8}$ and $a \to a+\frac{\tilde x^2}{32}$, $b \to b+\frac{\tilde x^2}{32}$ to obtain the above identity, but now $k$ is on 
a contour parallel to the real axis. Bringing back this contour to the real line, we obtain (\ref{id1}). 

The second identity reads, for $\tilde w>0$: 
\bea
\fl && ~~~~~    2 \int \frac{dk}{2 \pi} Ai(4 k^2 + a+b+ i k \tilde x) \frac{e^{2 i k (b-a)}}{i k + \tilde w} \label{id2} \\
\fl && ~~~~~~~~~~~~= \int_0^{+\infty} dr  ~ 2^{-\frac{1}{3}}  Ai(2^{\frac{1}{3}} (a+\frac{r}{4}+\frac{\tilde x^2}{32})) Ai(2^{\frac{1}{3}} (b-\frac{r}{4}+\frac{\tilde x^2}{32})) e^{\frac{\tilde x}{4} (b-a)- r (\frac{\tilde x}{8}+\tilde w)} \nn
\eea 
It is obtained by introducing an auxiliary variable writing $\frac{1}{i k + \tilde w} = \int_0^{+\infty} dr e^{- r( i k + \tilde w)}$. 
Then using the above identity (\ref{id1}) with $a \to a+\frac{r}{4}$, $b \to b-\frac{r}{4}$. \footnote{Alternatively one can first shift $i k \to i k +\tilde x/8$ in the l.h.s, which shows that the integral depends only on the STS invariant combination
$\tilde w + \frac{\tilde x}{8}$, and recover the r.h.s. However this is legitimate only if ${\rm sgn} \tilde w= {\rm sgn} (\tilde w + \frac{\tilde x}{8})$ 
otherwise we cross a pole and generate an additional pole contribution.} 

Note that if we set $\tilde w<0$ we can use instead $\frac{1}{i k + \tilde w} =-  \int_{-\infty}^0 dr e^{-r( i k + \tilde w)}$ hence
we obtain exactly the same integral with $\int_0^{+\infty} dr \to - \int_{-\infty}^0 dr$. Here we do not need $\tilde w<0$
however it is useful to consider the limit $\tilde w=0^+$ minus $\tilde w=0^-$, which picks up the residue of the
pole at $k=0$ leading to another useful identity:

\bea \label{id4} 
\fl && \int_{-\infty}^{+\infty} dr  ~ 2^{-4/3}  Ai(2^{\frac{1}{3}} (a+\frac{r}{4}+\frac{\tilde x^2}{32})) Ai(2^{\frac{1}{3}} (b-\frac{r}{4}+\frac{\tilde x^2}{32})) e^{- r \frac{\tilde x}{8}} = Ai(a+b) e^{\frac{\tilde x}{4} (a-b)}  
\eea 

Using (\ref{id1}) and (\ref{id2}) with $a=v_1+\frac{y+s}{2}$, $b=v_2+\frac{y+s}{2}$ 
we can rewrite the Kernel (\ref{K0}) {\it equivalently} \cite{footnote1} as:
\bea
\fl && K(v_1,v_2) = \int_{0}^{+\infty} dr \int \frac{dk}{2 \pi} dy ~ ( \theta(y) \delta(r) + \frac{1}{2} \delta(y) ) \\
\fl && \times 
2^{-1/3}  Ai\big(2^{\frac{1}{3}} (v_1 + \frac{y+s}{2} + \frac{r}{4} +\frac{\tilde x^2}{32}) )\big) Ai\big(2^{\frac{1}{3}} (v_2 + \frac{y+s}{2} - \frac{r}{4} +\frac{\tilde x^2}{32}) )\big) 
e^{- r (\tilde w+\frac{\tilde x}{8})} 
\eea
We now rescale $y  \to 2^{2/3} y $, $r  \to 2^{5/3} r $. Using the similarity transformation:
\bea
&& K(v_1,v_2) = 2^{\frac{1}{3}} \tilde K(2^{\frac{1}{3}} v_1,2^{\frac{1}{3}} v_2) 
\eea
and changing notations from $r \to y$ in the second integral, we obtain the equivalent form
for the Kernel:
\bea \label{result2}
\fl && g_\infty(s) = Prob( \xi < s) = Det[ I - {\cal K} ]   \quad , \quad {\cal K}(v_1,v_2) = \theta(v_1) \theta(v_2) \tilde K(v_1,v_2) \\
\fl && \tilde K(v_1,v_2) = K_{Ai}(v_1+\sigma,v_2+\sigma) + K^u_2(v_1+\sigma,v_2+\sigma) \quad , \quad \sigma = 2^{-2/3} (s + \frac{\tilde x^2}{16}) 
\eea 
in terms of the standard Airy Kernel (\ref{airyK}) and the Kernel:
\bea \label{result3} 
\fl && ~~ K^u_{2}(v_1,v_2)  =  \int_{y>0} dy Ai(v_1 + y) Ai(v_2 - y) e^{  2 y u} \quad , \quad u = - 2^{2/3} ( \tilde w + \frac{\tilde x}{8}) \\
\fl && ~~~~~~ = - \int_{y<0}  dy Ai(v_1 + y) Ai(v_2 - y) e^{ 2 y u } + 2^{-\frac{1}{3}} Ai(2^{-\frac{1}{3}} (v_1+v_2 - 2 u^2)) e^{- u(v_1-v_2)} \nn
\eea 
where we recall $\tilde w= \lambda w$, $x = \lambda^2 \tilde x$ and $\lambda=(t/4)^{1/3}$. The second form for 
$K_2$ is obtained by substituting $r \to 2^{5/3} y$, $a+\frac{\tilde x^2}{32} \to 2^{-1/3} v_1$, $b+\frac{\tilde x^2}{32} \to 2^{-1/3} v_2$ 
and $\frac{\tilde x}{8} \to - 2^{-2/3} u$ in (\ref{id4}).

\section{GUE and GOE limits and the connection to ${\cal A}_{2 \to 1}$} 
\label{sec:recover} 

It is easy to recover the GUE droplet limit by taking $\tilde w + \frac{\tilde x}{8} \to +\infty$ in (\ref{result2})-(\ref{result3}). 
One sees that $K_2$ vanishes in that limit, hence one recovers:
\bea
\fl && ~~~~~ g_\infty(s) =  Prob( \xi < s)  \to F_2(\sigma=2^{-2/3} (s + \frac{\tilde x^2}{16}) ) ~~ \Leftrightarrow ~~~ \lambda \xi = \chi_2 t^{1/3} - \frac{x^2}{4 t}
\eea 
which coincides with the known result (\ref{h2}). Note that it is in fact a double limit, where 
we keep the combination $s+ \frac{\tilde x^2}{16}$ finite, to account for the average profile $- \frac{x^2}{4 t}$ of the droplet solution. 

To recover the GOE flat limit, we need to consider $\tilde w + \frac{\tilde x}{8} \to - \infty$ (while
keeping $\tilde w>0$, a purely technical restriction). This time, since the average profile is flat we keep
$s$ finite. Hence $K_{Ai}(v_1+\sigma,v_2+\sigma)$ vanishes in that limit, and so does the first term
in the second expression for $K_2$. Hence we are left with only the second piece of $K_2$,
more precisely:
\bea
{\cal K}(v_1,v_2) \simeq_{\tilde{x} \to - \infty}  2^{-\frac{1}{3}} Ai(2^{-\frac{1}{3}} (v_1+v_2 + 2 \sigma - 2 u^2)) \theta(v_1) \theta(v_2)
\eea 
where we have again discarded the factor $e^{- \tau(v_1-v_2)}$ which is immaterial in calculating the Fredholm determinant. Performing
the change $v_1 \to 2^{1/3} v_1$, $v_2 \to 2^{1/3} v_2$ and comparing with (\ref{defxi}) we obtain
\bea
\fl && ~~~~~ g_\infty(s) =  Prob( \xi < s)  \to F_1(s - 4 \tilde w^2 - \tilde w \tilde x) ~~ \Leftrightarrow ~~~ \lambda \xi = \lambda \chi_1 + w x + t w ^2
\eea 
hence we recover the GOE Tracy Widom distribution, up to a shift equal to the average profile, i.e. the solution of the KPZ equation
in the absence of noise with initial condition $w x$.

One can now compare (\ref{result2})-(\ref{result3}) with the result of the Appendix A of \cite{BorodinAiry2to1}. The Kernel there is 
identical to ours provided we identify:
\bea
&& \!\!\!\!\!\!\!\!\! 2^{-2/3} \xi = - 2^{-2/3} \frac{\tilde x^2}{16} + {\cal A}_{2 \to 1}(u) + \max(0,u)^2 \quad , \quad u = - 2^{2/3} ( \tilde w + \frac{\tilde x}{8}) 
\eea 
This can also be written as ($\xi$ being defined in (\ref{defxi})):
\bea
\lambda \xi = - \frac{x^2}{4 t} + t^{1/3} (  {\cal A}_{2 \to 1}(u) - \min(0,u)^2 + u^2 ) 
\eea 
which is exactly (in rescaled variables, and for $z=0$) the prediction of Eq. (\ref{resa21}) obtained there from 
the argument of dominance of the maximum in the large time limit. Hence
we recover quite precisely for the KPZ equation, from the RBA calculation, the transition process which was obtained
in the context of the TASEP. This shows the desired universality. 

\newpage

\section{Generalization: maxima of the ${\cal A}_{2 \to 1}$ transition process}
\label{sec:general} 

\subsection{definitions and relations} 

\begin{figure}[htpb]
  \centering
  \includegraphics[width=0.4\textwidth]{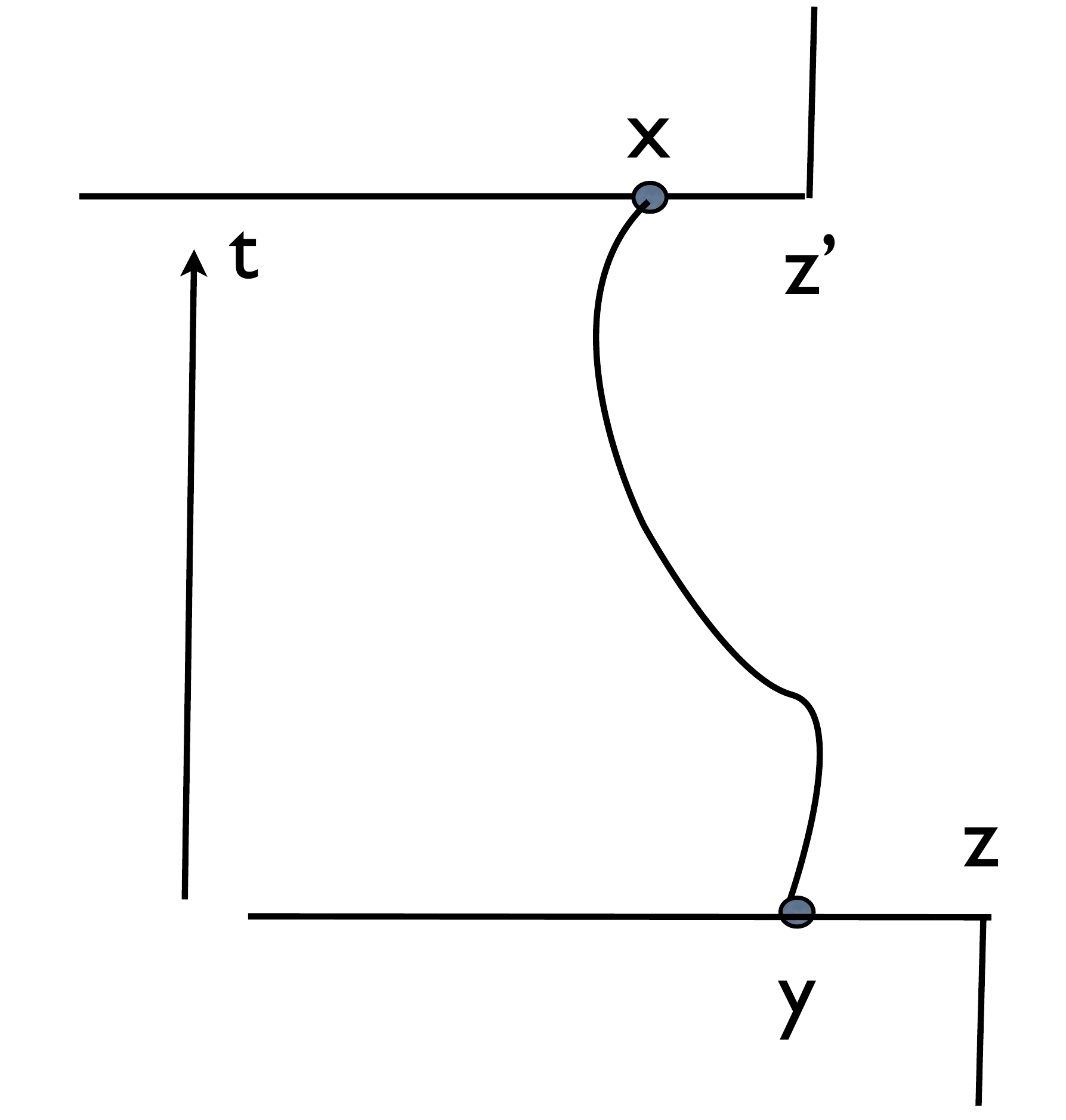}
  \caption{LL geometry: each endpoint of the DP is free on the left half-spaces $x<z'$ and $y<z$, respectively,
  with in addition exponential weights which amount (for $w,w'>0$) to tilt the half-lines, $e^{w' x}$ (tilt upward at $-\infty$)
  and $e^{w y}$ (tilt downward at $-\infty$).}
\label{fig:fig2}

\end{figure}

\begin{figure}[htpb]
  \centering
  \includegraphics[width=0.4\textwidth]{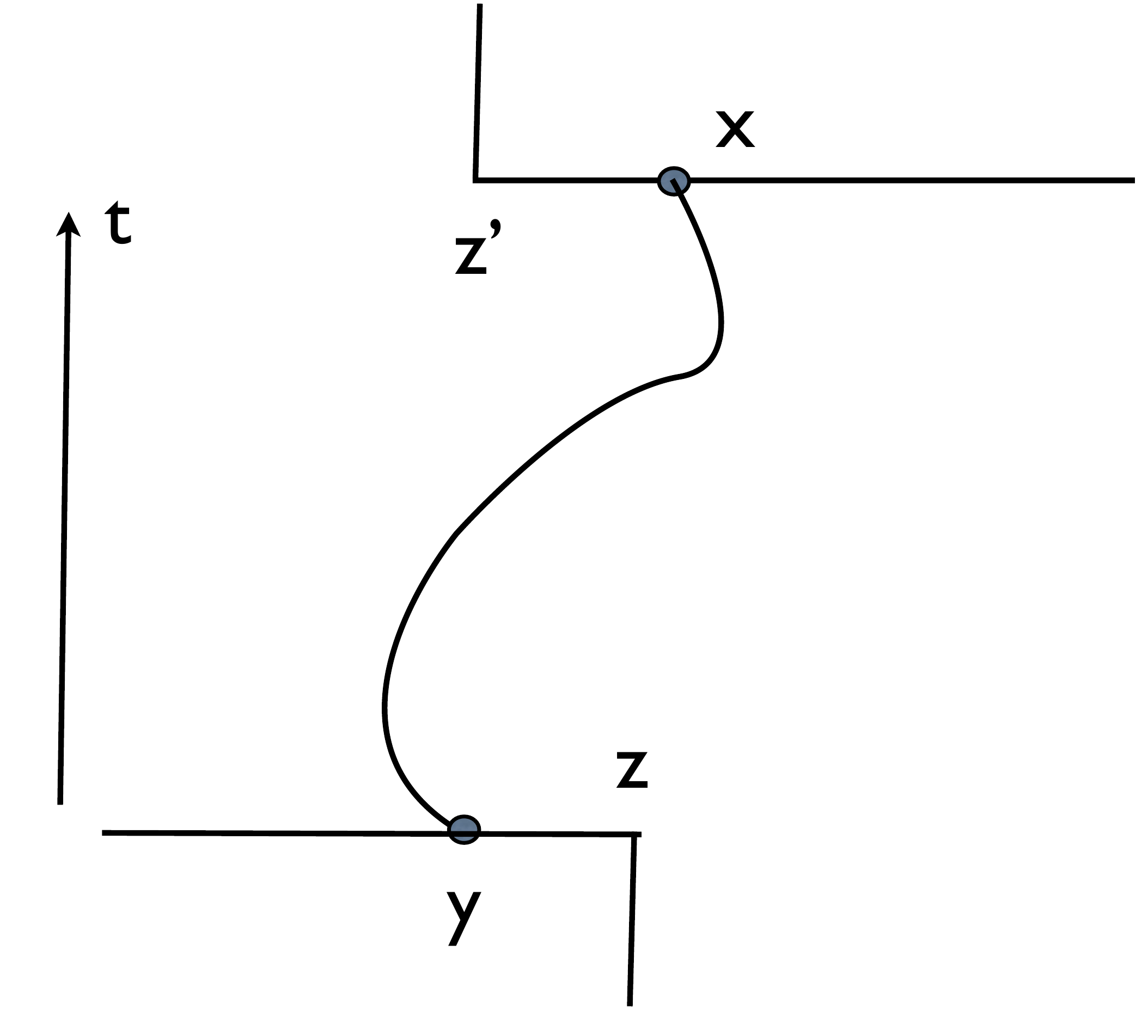}
  \caption{LR geometry: each endpoint of the DP is free on the a half-spaces $x>z'$ (right half-space) and $y<z$ (left half-space), respectively,
  with in addition exponential weights which amount (for $w,w'>0$) to tilt the half-lines, $e^{-w' x}$ (tilt upward at $+\infty$)
  and $e^{w y}$ (tilt downward at $-\infty$).}
\label{fig:fig3}
\end{figure}

It turns out that the Bethe Ansatz formula (\ref{sumf}) is easily generalized to study the partition
sums of a DP where each of the two endpoints is free to explore its own half-space.
This in turn gives interesting information on the extremal properties of the
transition process ${\cal A}_{2 \to 1}$ (minus a half-parabola). There are clearly two cases, either the
two half-spaces are on the same side, or on opposite sides, see Fig. \ref{fig:fig2} and Fig. \ref{fig:fig3}. 
We thus now define:
\bea
\fl && ~~~~~ h^{LL}_{zwz'w'}(t) : = \ln \int_{x<z'} dx \int_{y < z} dy e^{h(x,t|y,0)+ w y + w' x} =  \ln \int_{x < z'} dx e^{h^L_{wz}(x,t) + w' x}  \\
\fl &&  ~~~~~ h^{LR}_{zwz'w'}(t) : = \ln \int_{x > z'} dx \int_{y < z} dy e^{h(x,t|y,0)+ w y - w' x} = \ln \int_{x > z'} dx e^{h^L_{wz}(x,t) - w' x} 
\eea
Note that $h^{LR}$ is well defined for $w=w'=0$, since, as can be seen in Fig. \ref{fig:fig3} the 
line tension of the polymer (i.e. the diffusion kernel) makes all integrals convergent. On the
other hand $h^{LL}$ is finite only for $w,w'>0$ since there is nothing to prevent the
polymer to be arbitrary far to the left, see Fig. \ref{fig:fig2}. 

Again we can expect that in the large $t$ limit these integrals are dominated by their maximum. 
One thus expects, setting $w=w'=0$ that:
\bea
&& h^{LR}_{zz'}(t) = \ln \int_{x > z'} dx e^{h^L_{z}(x,t)}  \to_{t \to + \infty} \max_{x>z'} h^L_z(x,t) \\
&& \simeq t^{1/3} \max_{x>z'} [ {\cal A}_{2 \to 1}(v) - \min(0,v)^2 ]_{v=(z-x)/(2 t^{2/3})} \\
&& = t^{1/3} \max_{v < u=\frac{z-z'}{2 t^{2/3}}} [ {\cal A}_{2 \to 1}(v) - \min(0,v)^2  ] \label{max5} 
\eea 
where in the second line we have used (\ref{hLAiry}). Because of the term $- \min(0,v)^2$ this
optimization problem is well defined for $v \to -\infty$, the GUE side. Optimization
on the complementary interval $[u,+\infty[$ (the GOE side) is clearly divergent. 

Keeping now $w,w'>0$ we can also write:
\bea
\fl && ~~~~~~~~~ h^{LL}_{zwz'w'}(t) = \ln \int_{x < z'} dx e^{h^L_{z}(x,t) + w' x}  \to_{t \to + \infty} \max_{x<z'} (h^L_{zw}(x,t) + w' x) \\
\fl && ~~~~ \simeq t w^2 +  \max_{x<z'} \bigg( (w+w') x + t^{1/3} [{\cal A}_{2 \to 1}(v) - \min(0,v)^2 ]_{v=- t^{1/3} w + \frac{z-x}{2 t^{2/3}}} \bigg) \\
\fl && ~~~~ = - t w^2 - 2 t w w' + (w+w') z  \nn \\
\fl && ~~~~~~~~~~~~~~ + t^{1/3} \max_{v> u= \frac{z-z'-2 t w}{2 t^{2/3}}} [ {\cal A}_{2 \to 1}(v) - \min(0,v)^2 - 2 v \times 2^{2/3} (\tilde w+\tilde w') ]
\label{maxAiry21new} 
\eea 
where in the second line we have used (\ref{resa21}) and we require $w+w'>0$ for the problem to be well defined. We have defined
the scaled variables $\tilde w= \lambda w$ and $\tilde w'=\lambda w'$ hence $2^{2/3} (\tilde w+\tilde w')=t^{1/3} (w+w')$. 
Note the symmetry property which arises from the above definitions: $h^{LL}_{zwz'w'}(t) \equiv h^{LL}_{z'w'zw}(t)$ which
implies $\max_{x<z} (h^L_{z'w'}(x,t) + w x) \equiv \max_{x<z'} (h^L_{zw}(x,t)+ w' x)$, hence (\ref{maxAiry21new}) must also
be symmetric w.r.t to the exchange of $(z,w)$ with $(z',w')$. This can be checked in the
absence of noise where the solution, in the infinite time limit, reads (this amounts to set ${\cal A}_{2 \to 1}(v) \to 0$ in
 (\ref{maxAiry21new})):
\bea
\fl && h^{LL0}_{zwz'w'}(t)  = \max_{x<z', y<z} (- \frac{(x-y)^2}{4 t} + w y + w' x) = \theta(z'-z>2 w' t) [ t w'^2 + (w+w') z ]  \\
\fl && + \theta(2 w t< z'-z<2 w' t) [ w z + w' z' - \frac{(z-z')^2}{4 t}] + \theta(z'-z<-2 w t) [ t w^2 + (w+w') z' ] \nn 
\eea
which is complicated but clearly symmetric in the exchange of $(z,w)$ with $(z',w')$. The fact that
such a symmetry holds also (in law) for the noisy case is not quite easy to guess from
just looking at (\ref{maxAiry21new}). However it must be correct, and we do indeed find
that our result below satisfies it. 


\subsection{quantum mechanics} 

Let us now express the moments of the partition sums:
\bea
&& Z^{LL}(t) = Z^{LL}_{zwz'w'}(t) = e^{h^{LL}_{zwz'w'}(t)} \\
&& Z^{LR}(t) = Z^{LR}_{zwz'w'}(t) = e^{h^{LR}_{zwz'w'}(t)}
\eea 
using the eigenstates $\mu = \{\lambda_1,..\lambda_n\}$ of the Lieb-Liniger model. Generalizing formula (\ref{sumf})  we see that:
\bea
&& \overline{(Z^{LL})^n} = 
\sum_\mu ( I_\mu^L(w') )^*  e^{ (n w'- i \sum_\alpha \lambda^*_\alpha) z'}  
\frac{e^{-t E_\mu} }{||\mu ||^2} e^{ (n w+i \sum_\alpha \lambda_\alpha) z}  I_\mu^L(w)  \\
&& \overline{(Z^{LR})^n} = 
\sum_\mu ( I_\mu^R(w') )^*  e^{ (- n w'- i \sum_\alpha \lambda^*_\alpha) z'}  
\frac{e^{-t E_\mu} }{||\mu ||^2} e^{ (n w+i \sum_\alpha \lambda_\alpha) z}  I_\mu^L(w)  
\eea
where we have defined:
\bea
&& I_\mu^L(w) = \prod_{j=1}^n \int_{-\infty}^0 dy_j e^{w y_j}  \Psi_\mu(y_1\dots y_n)  \\
&& I_\mu^R(w') = \prod_{j=1}^n \int^{+\infty}_{0} dx_j e^{- w' x_j} \Psi_\mu(x_1 \dots x_n) = I_{-\mu}^L(w')
\eea
where $-\mu$ denotes the state with reversed rapidities $-\mu =  \{- \lambda_1,..- \lambda_n\}$ and we use $\Psi_{\mu}(-x_1,..-x_n)=\Psi_{-\mu}(x_1,..x_n)$, as can be seen from (\ref{def1}). 

Up to now the formula are general. Let us now consider the limit $L \to +\infty$ and insert the string states. 
Let us define:
\bea \label{shift}
&& \ln Z^{LL} = w z + w' z' + \ln \tilde Z^{LL} \quad , \quad  \ln Z^{LR} = w z - w' z' + \ln \tilde Z^{LR} 
\eea 
From the results of \cite{we-flat,we-flatlong} (Eq. (76)), and Section \ref{sec:start} (see notations there) we have:
\bea
&& I_\mu^L(w) = n! (-2)^n S^w[k,m] D^w[k,m] \\
&& I_\mu^R(w') = n! (-2)^n S^{w'}[-k,m] D^{w'}[- k,m] = I_\mu^L(w')^*
\eea
and we recall the property 
\bea 
&&  (S^w_{m,k})^* = S^w_{m,- k}  \quad , \quad (D^w_{m_1,k_1,m_2,k_2})^*=D^w_{m_1,-k_1,m_2,-k_2}
\eea
from the definitions (\ref{sw}), (\ref{dw}). 

The moments thus read:
\bea
\fl && \overline{(\tilde Z^{\epsilon})^n} = \sum_{n_s=1}^n \frac{n! (-4)^n}{n_s!} \sum_{(m_1,..m_{n_s})_n} 
\Phi[k,m] S^w[k,m] S^{w'}[ \epsilon k,m] D^w[k,m] D^{w'}[ \epsilon k,m] \\
\fl && ~~~~~ ~~~~~ ~~~~~ ~~~~~ \times \prod_{j=1}^{n_s} \frac{1}{m_j} e^{m_j^3 \frac{t}{12}- m_j k_j^2 t  - i (z'-z) m_j k_j }
\eea
where from now on we denote:
\bea
LL ~~ \Leftrightarrow  ~~ \epsilon = -1 \quad , \quad LR ~~ \Leftrightarrow  ~~ \epsilon = +1
\eea 

We then define the generating functions:
\bea
g_\lambda^{LL}(s) = \sum_{n=0}^{+\infty} \frac{1}{n!} e^{- \lambda n s} \overline{(\tilde Z^{LL})^n} 
\eea 
and similarly for $g_\lambda^{LR}(s)$. It is easy to see that $g_\lambda^{\epsilon}(s)$ is then given exactly by the same formula as $g_\lambda(s)$ in (\ref{start})
with the substitution \cite{notes} 
\bea
&& S^w_{m_j,k_j}  \to  (-2)^{m_j}  S^w_{m_j,k_j}   S^{w'}_{m_j, \epsilon k_j} \quad , \quad x \to z'-z \\
&& D^w_{m_i,k_i,m_j,k_j} \to D^w_{m_i,k_i,m_j,k_j}  D^{w'}_{m_i, \epsilon k_i,m_j, \epsilon k_j}  
\eea 

\subsection{large time limit and first form for the Kernel} 

We now consider the large time limit and again set all factors $D^{w}$, $D^{w'}$ to unity. 
We define again the scaled variables $\tilde z=z/\lambda^2$, $\tilde z'=z'/\lambda^2$, $\tilde w= \lambda w$, 
$\tilde w'= \lambda w'$. Performing exactly 
the same steps as in Sections \ref{sec:rescaling} and \ref{sec:largetime}, and using the formula (\ref{sw}) 
we arrive at the following expressions in terms of Fredholm determinants:
\bea
 \! \! \! \!   \! \! \! \!  g_\lambda^{\epsilon}(s) \to Det[ I - {\cal M}] \quad , \quad {\cal M}(v_1,v_2) = \theta(v_1) \theta(v_2) M(v_1,v_2) 
\eea 
with the same Kernel M as in ({\ref{M}) (with $\tilde x=\tilde z-\tilde z'$) with:
\bea
&& \! \! \! \!  \! \! \! \!  \phi_\lambda(k,y)  \to \phi^{\epsilon}_\lambda(k,y) =
\sum_{m=1}^\infty \frac{(-1)^{m} 2^{2m+1} \Gamma(\frac{2 i k + 2 \tilde w}{\lambda}) \Gamma(\frac{2 i k \epsilon + 2 \tilde w'}{\lambda})}{\Gamma(\frac{2 i k + 2 \tilde w}{\lambda} + m)\Gamma(\frac{2 i k \epsilon + 2 \tilde w'}{\lambda} + m)} e^{\lambda m y} 
\eea 
We again use the Mellin Barnes identity (\ref{MB}) and rescale $s_j \to s_j/\lambda$, to obtain, in the limit $\lambda \to + \infty$:
\bea
&& \! \! \! \! \! \! \! \! \! \! \! \! \phi^{\epsilon}_{+\infty}(k,y) =  \int_{C}  \frac{- ds}{2 i \pi s} \frac{1}{2}  \frac{(2 i k + 2 \tilde w + s)(2 i k \epsilon + 2 \tilde w' + s)}{(i k + \tilde w)( i k \epsilon + \tilde w')} e^{s y} \\
&& \! \! \! \! \! \! \! \! \! \! \! \! = \int_{C}  \frac{- ds}{2 i \pi s}  [2 +  \frac{s}{i k + \tilde w} +  \frac{s}{ i k \epsilon + \tilde w'} + \frac{1}{2} \frac{s^2}{(i k + \tilde w)( i k \epsilon + \tilde w')} ] e^{s y} \\
&& \! \! \! \! \! \! \! \! \! \! \! \! = -  2 \theta(y) -  [ \frac{1}{i k + \tilde w} + \frac{1}{i k \epsilon + \tilde w'} ] \delta(y) - \frac{1}{2}  \frac{1}{(i k + \tilde w)(i k \epsilon + \tilde w')} \delta'(y) 
\eea 
%

Hence we finally obtain that: 
\bea \label{defxi2} 
h^\epsilon_{zwz'w'}(t) = v_0 t + w x - \epsilon w' x' + \lambda \xi \quad , \quad \lambda = (t/4)^{1/3} 
\eea 
with, as $t \to \infty$:
\bea \label{fd2} 
\fl && ~~~~~ Prob(\xi < s) = g_\infty^{\epsilon}(s) = Det[ I - {\cal K}^{\epsilon} ] \quad , \quad {\cal K}^{\epsilon}(v_1,v_2) = \theta(v_1) \theta(v_2) K^{\epsilon}(v_1,v_2) 
\eea 
with the Kernels:
\bea  \label{KLL} 
\fl && K^{\epsilon}(v_1,v_2) =   \int \frac{dk}{2 \pi} dy \bigg[ 2 \theta(y) +  ( \frac{1}{i k + \tilde w} + \frac{1}{i k \epsilon + \tilde w'} ) \delta(y) 
- \frac{1}{2}  \frac{1}{(i k + \tilde w)(i k \epsilon + \tilde w')} \delta(y) \partial_y \bigg] \nn \\
\fl && ~~~~~~~~~~~~~~~~~~~~~~~~~~~~~~~~~~~~
 \times Ai(y + 4 k^2 + i k (\tilde z' - \tilde z ) + v_1 + v_2 + s) e^{-2 i k (v_1-v_2)} 
\eea 
where $\tilde w, \tilde w'>0$ and we recall that $\epsilon=-1$ for $LL$ and $\epsilon=+1$ for $LR$.
The STS symmetry is again recovered from the shift $i k \to i k + \frac{\tilde z'-\tilde z}{8}$ followed 
by the shift back to the real axis, and it shows that all the dependence in $z,z',w,w'$ has the form \cite{footnote1}:
\bea \label{sym1} 
g_\infty^\epsilon(s) = \tilde g^\epsilon_\infty(s + \frac{(\tilde z'-\tilde z)^2}{16} , \tilde w + \frac{\tilde z'-\tilde z}{8} , \tilde w' + \epsilon \frac{\tilde z'-\tilde z}{8})
\eea 
which is a symmetric function of its last two arguments (obvious for $\epsilon=+1$ (LR) and using $k \to -k$ followed by the transposition of the Kernel for $\epsilon=-1$ (LL)). The function $\tilde g^\epsilon_\infty$ 
has the same form (\ref{fd2}) with the same Kernel $K$ where $\tilde z - \tilde z'$ is set to 0. 

One easily checks some limits. Taking both $\tilde w, \tilde w' \to + \infty$ in (\ref{KLL}) one recovers the
GUE Kernel. The same is thus true if one takes $\tilde z'-\tilde z \to + \infty$. If one takes
$\tilde w' \to + \infty$ at fixed $w$ one recovers the half-flat Kernel (\ref{result2})-(\ref{result3}) as
expected.

\subsection{second form for the Kernel} 

We again transform these Kernels to another equivalent form. Details are given in \ref{sec:manip}. 
Our final result is that (\ref{defxi2}) holds with:
\bea \label{result5}
\fl && ~~~~~~ g^\epsilon_\infty(s) = Prob( \xi < s) = Det[ I - {\cal K}^\epsilon ]   \quad , \quad {\cal K}^\epsilon(v_1,v_2) = \theta(v_1) \theta(v_2) \tilde K^\epsilon(v_1,v_2) 
\eea
with $\epsilon=-1$ for $LL$, $\epsilon=+1$ for $LR$  and  the Kernel \cite{footnote1}:
\bea \label{result6} 
\fl &&  \tilde K^{\epsilon}(v_1,v_2) = K_{Ai}(v_1+\sigma,v_2+ \sigma)  + \bar K^{\epsilon}(v_1+\sigma,v_2+\sigma)  \quad , \quad \sigma = 2^{-2/3} ( s + \frac{(\tilde z'-\tilde z)^2}{16}) \nn
\eea
where $K_{Ai}$ is the Airy Kernel (\ref{airyK}) and:
\bea
\fl && ~~~ \bar K^{LL}(v_1,v_2) = K_2^u(v_1,v_2) + K_2^{u'}(v_2,v_1) + K_3^{LL}(v_1,v_2) \\
\fl && ~~~ \bar K^{LR}(v_1,v_2) = K_2^u(v_1,v_2) + K_2^{u'}(v_1,v_2) + K_3^{LR}(v_1,v_2) \\
\fl && ~~~~~~~~~~~ u = - 2^{2/3} ( \tilde w + \frac{{\tilde z}'-\tilde z}{8})  \quad , \quad u' = - 2^{2/3} ( \tilde w' + \epsilon \frac{{\tilde z}'-\tilde z}{8})
\eea
in terms of the Kernel $K_2^u$ defined in (\ref{result3}) and in terms of a third Kernel:
\bea
\fl && K_3^{LL}(v_1,v_2) = (\partial_{v_1}+\partial_{v_2}) \int_0^{+\infty} dr \frac{Ai(v_1+r)Ai(v_2-r) e^{2 r u} + Ai(v_1-r)Ai(v_2+r) e^{2 r u'}}{2(u+u')}   \nn \\
\fl && K_3^{LR}(v_1,v_2) = (\partial_{v_1}+\partial_{v_2}) \int_0^{+\infty} dr Ai(v_1+r)Ai(v_2-r) \frac{e^{2 r u} - e^{2 r u'}}{2(u'-u)}  \label{last}
\eea 
We note that $\tilde K^{LR}$ is invariant in the exchange of $u$ and $u'$, while $\tilde K^{LL}$
is changed in its transpose \cite{footnote2} hence we recover the symmetry of the function $\tilde g^\epsilon_{\infty}$ in
Eq. (\ref{sym1}). 

\subsection{maximum of the transition process} 

As one example of application, consider $g^{LR}(s)$ and set $w=w'=0$ hence $u=u'$. Putting together (\ref{max5}) together with our above result (\ref{result5})-(\ref{last}) and the definition of $\xi$ (\ref{defxi2}) we obtain that:
\bea
\fl && ~~~~~~~  t^{1/3} \max_{v < u} [ {\cal A}_{2 \to 1}(v) - \min(0,v)^2  ] = t^{1/3} \chi^u - \frac{(z-z')^2}{4 t} = t^{1/3} (\chi^u - u^2) 
\eea 
where $\chi^u$ is distributed according to:
\bea
\fl && ~~~~~~~ Prob( \chi^u < S) = {\rm Det}[I - {\cal P}] \quad , \quad {\cal P}(v_1,v_2) = \theta(v_1) \theta(v_2) P(v_1+S,v_2+S) \\
\fl && ~~~~~~~ P(v_1,v_2) = K_{Ai}(v_1,v_2) + ( 2 - \frac{1}{2} (\partial_{v_1} + \partial_{v_2}) \partial_u ) K_2^u(v_1,v_2) 
\eea 
More results concerning the maximum of ${\cal A}_{2 \to 1}(v) - \min(0,v)^2 - b v$ on intervals either
$v \in [u,+\infty[$ (RR) or $v \in ]-\infty,u]$ are also easily extracted from the above formula
(in view of e.g. (\ref{maxAiry21new})).

\section{Conclusion}

To conclude we have shown how the replica Bethe Ansatz allows to find the large time limit of the PDF of the
height field of the continuum KPZ equation for the half-flat initial condition. This PDF takes the form a of Fredholm determinant.
We show that its Kernel can be transformed to the one found for the TASEP with the corresponding initial condition,
a manifestation of KPZ universality. Being confident that the procedure for taking the large time limit gives the
correct answer, although at this stage not fully justifiable, we obtain the one point PDF for the problem of
a directed polymer with each endpoint on its own half-space. It also takes the form of Fredholm determinants
with new Kernels that we display and analyze. These are related to the extremal statistics of the
half-flat initial condition. It seems that it should now be doable, by taking a similar limit, to study 
the many point problem and recover the complete Airy$_{2 \to 1}$ process. 

It would also be interesting to test numerically the present results, or in liquid crystal experiments.
They also have consequences 
for the conductance $g$ of disordered 2D conductors deep in
the Anderson localized regime, specifically for the conductance 
from a point lead $(x,t)$ to an extended lead $(0,y)$ $y \in [0,L[$.
Extending the results of Ref. \cite{Ortuno} we surmise that the one point distribution of $\ln g$, 
scaled by $t^{1/3}$ should exhibit the
GUE to GOE universal crossover distribution. Similarly our LR (and LL)
results can in principle be tested for two parallel half-line leads (eventually tilted by small angles $\sim w,w'$).
This could provide a rather detailed test of the conjectured relation
between the positive weight DP problem and the Anderson problem, which, at present, is 
not based on any exactly solvable model. 

\bigskip

{\it Acknowledgments:}
I am grateful to J. Quastel for suggesting the first part of this calculation, and informing me of work in preparation
on the flat and half-flat initial condition \cite{Quastelflat} . I thank
P. Calabrese for discussions and collaborations closely related to many aspects of this work. I thank A. Borodin for useful discussions. 

\appendix

\section{Large time limit}
\label{sec:large} 

The rationale to set the $D^w$ factor to unity in the large time limit is as follows. 
Assume one can use the Mellin Barnes identity (\ref{MB}) on the
starting formula (\ref{2}) and write:
\bea
&& \fl 
 Z(n_s,s) =  
 \prod_{j=1}^{n_s}  \int_{C_j}   \frac{- ds_j}{2 i \pi \sin \pi s_j}  2^{s_j+1}  \int \frac{dk_j}{2 \pi} \int dy_j  
 Ai(y_j + 4 k_j^2 + i k_j \tilde x + s)
 \frac{ \Gamma(\frac{2 i k_j + 2 \tilde w}{\lambda}) e^{\lambda s_j y}}{\Gamma(\frac{2 i k_j + 2 \tilde w}{\lambda} + s_j)} 
  \nn  \\
&& \fl \times  \prod_{1 \leq i < j \leq n_s} D^w_{s_i,k_i/\lambda,s_j,k_j/\lambda}  \times det[ \frac{1}{ 2 i (k_i - k_j) + \lambda s_i + \lambda s_j }]_{n_s \times n_s} \label{new} 
\eea
where $C_j=a+i r_j$, $0<a<1$. Then rescaling $s_j \to s_j/\lambda$ and take the large $\lambda$ limit
one finds that indeed $D^w_{s_i,k_i/\lambda,s_j,k_j/\lambda} \to 1$ from the definition 
(\ref{dw}). Pursuing the calculation then immediately leads to the same formula as in 
Section \ref{sec:largetime}, as the $s_j$ integral then decouple.

This is the type of argument which was used in Ref. \cite{dotsenkoGOE,dotsenko2pt,dotsenkoEndpoint,dotsenko2times}.
At present its proper justification escapes us, however. One condition for the Mellin Barnes identity (\ref{MB})
to hold is that $f(s_j)$ has no poles for $Re s_j>a$. One easily sees that the determinant is not a
problem as its poles for $s_i+s_j$ live on the imaginary axis. However examination of the formula 
(\ref{dw}) for $D^w_{s_i,k_i/\lambda,s_j,k_j/\lambda}$ (if we take it as the proper analytical continuation for complex
$m_j$)
shows that at fixed $k_j$ it has numerous poles
where some of the $Re s_j>a$.

Given that we know that setting $D^w \to 1$ does give the correct answer in a number of cases,
including in the present study, it is quite likely that there is indeed an integral representation either 
identical or similar to the above one. One way to check would be to actually enumerate the
additional poles, add their contributions and show that they indeed vanish for infinite $\lambda$.
Another check would be to see whether the finite time solution of \cite{we-flat,we-flatlong} can be
also retrieved from (\ref{new}) or a modification thereof. Finally, other analytical continuations in the $m_j$ than (\ref{dw}) may be searched for.
These go beyond the present study.
It is likely that a better understanding of why and how this limit works will come from the other routes, as limits from the
ASEP, $q$-TASEP or the O'Connor semi-discrete polymer, which do produce better controled
nested contour integral formula (see e.g. \cite{BorodinMacdo,Quastelflat}). 

\section{Kernel manipulations}
\label{sec:manip}

Let us start with the Kernel in (\ref{KLL}).
Writing in the last term $\frac{1}{(i k + \tilde w)(i k \epsilon + \tilde w')} = \frac{1}{\tilde w'-\epsilon \tilde w} (\frac{1}{i k + \tilde w} - \frac{\epsilon}{i k \epsilon + \tilde w'})$ using 
(\ref{id2}) with $a=v_1+\frac{y+s}{2}$, $b=v_2+\frac{y+s}{2}$  and rescaling $y  \to 2^{2/3} y $, $r  \to 2^{5/3} r $ we find:
\bea
&& K^\epsilon(v_1,v_2) = 2^{1/3} \tilde K^\epsilon(2^{1/3} v_1,2^{1/3} v_2) 
\eea
Hence we obtain Eqs. (\ref{result5}) and (\ref{result6}) in the text with:
\bea
\fl && \bar K^{\epsilon}(v_1,v_2) = \int_{0}^{+\infty} dr \int \frac{dk}{2 \pi} dy  \delta(y) (1- \frac{1}{2\times 2^{2/3}(\tilde w' - \epsilon \tilde w)} \partial_y) ) \\
\fl && \times 
Ai\big(v_1 + y + r \big) Ai\big(v_2 + y - r \big) 
e^{- 2 r 2^{2/3} (\tilde w+\frac{\tilde z'-\tilde z}{8})} \\
\fl && + \int_{0}^{+\infty} dr \int \frac{dk}{2 \pi} dy  \delta(y) (1+ \frac{\epsilon}{2\times 2^{2/3} (\tilde w' - \epsilon \tilde w)} \partial_y) ) \\
\fl && \times 
Ai\big(v_1 + y + \epsilon r \big) Ai\big(v_2 + y - \epsilon r \big) 
e^{- 2 r 2^{2/3} (\tilde w'+ \epsilon \frac{\tilde z'-\tilde z}{8})} 
\eea
This then leads to the form given in the text.


\begin{thebibliography}{32}
\bibitem{footnote1} 
The global multiplicative factor $e^{\frac{\tilde x}{4} (v_2-v_1)}$ eliminates in the calculation of the
Fredholm determinant hence we set it to unity.
\bibitem{footnote2}
Note that in the limit $u+u' = - 2^{2/3} (\tilde w+\tilde w') \to 0^-$ the Kernel $\tilde K^{LL}(v_1,v_2)$ diverges as $\sim \frac{2^{-2/3}}{u+u'}  Ai'(2^{-1/3} (v_1+v_2))$  using (\ref{result3}). This divergence was expected, as discussed above, the factor $1/(w+w')^{n_s}$ arising from
integration over the $n_s$ center of mass of strings. Whether some physics can be extracted from it remains to be explored.
Indeed, one expects a divergence $\overline{Z^n} \sim 1/(w+w')^n$ hence that the term $n_s=n$ dominates in the sum which
may require a different analysis.
\end{thebibliography}
\end{document}